\documentclass{iau}
\usepackage{graphicx}

\title[SMBBHs and TDEs] 
{Compact object mergers: Observations of supermassive
 binary black holes and stellar tidal disruption events}

\author[S. Komossa, J.A. Zensus]   
{S. Komossa$^{1}$,
J.A. Zensus$^1$
 }

\affiliation
{$^1$ Max-Planck-Institut f{\"u}r Radioastronomie, Auf dem H{\"u}gel 69, 53121 Bonn, Germany
}

\pubyear{2015}
\volume{312}  
\pagerange{119--126}
\jname{Star clusters and black holes in galaxies across cosmic time}
\editors{R. Spurzem, B.D. Editor \& C.E. Editor, eds.}
\begin{document}

\maketitle

\begin{abstract}
The capture and disruption of stars by supermassive black holes (SMBHs), and
the formation and coalescence of binaries, are inevitable consequences
of the presence of SMBHs at the cores of galaxies.
Pairs of active galactic nuclei (AGN) and binary SMBHs
are important stages in the evolution of galaxy mergers, and an intense search
for these systems is currently ongoing. In the early and advanced
stages of galaxy merging, observations of the triggering of accretion
onto one or both BHs inform us about feedback processes and BH
growth. Identification of the compact binary SMBHs at parsec and sub-parsec scales
provides us with important constraints on the interaction processes that
govern the shrinkage of the binary beyond the ``final parsec''. Coalescing
binary SMBHs are among the most powerful sources of gravitational waves (GWs)
in the universe. 
Stellar tidal disruption events (TDEs) appear as luminous, transient, accretion flares
when part of the stellar material is accreted by the SMBH. About 30 events
have been identified by multi-wavelength observations by now, and they 
will be detected in the thousands in future ground-based or space-based
transient surveys. The study of TDEs 
provides us with a variety of new astrophysical tools and applications,
related to fundamental physics or astrophysics. 
Here, we provide a review of the current status of observations of 
SMBH pairs and binaries, and TDEs, and discuss astrophysical implications.
\end{abstract}

\firstsection 
\section{Introduction: galaxy mergers and supermassive binary black holes}

Galaxies have merged frequently with each other throughout the history of the universe.
Galaxy mergers trigger quasars, are the sites of major black hole growth, and are believed to
fix the scaling relations between SMBH mass and host properties -- either by 
merging repeatedly with each other, and/or by feedback processes following the onset of
accretion onto one or both black holes.  
If both galaxies harbor SMBHs at their centers, these two will ultimately form a bound pair.
Coalescing supermassive binary black holes (SMBBHs) are among 
the most powerful emitters of gravitational
waves in the universe. The subsequent gravitational wave recoil of the newly formed
single black hole will, in rare cases, lead to kick velocities exceeding the host's escape 
velocity, and astrophysical consequences of this phenomenon are now being explored.  


\begin{figure}[t]
\begin{center}
 \includegraphics[width=10cm]{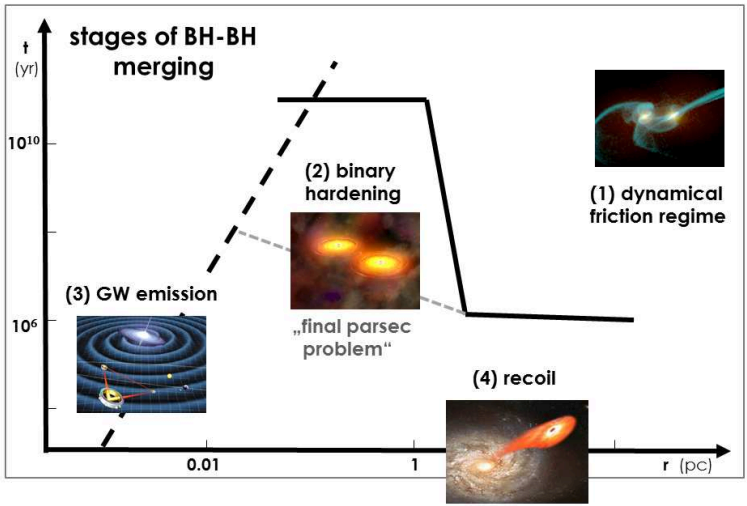}
 \caption{Stages of the evolution of SMBH pairs in the course of galaxy 
merging (following Begelman et al. 1980).
}
   \label{fig1}
\end{center}
\end{figure}

Galaxy and SMBH mergers evolve in several stages (e.g., 
Begelman et al. 1980, Roos 1981, Merritt \& Milosavljevic 2005,
Colpi 2014; our Fig. 1).
During the first stage, merging of the two galaxies is dominated by dynamical friction. 
At separations on the order of parsecs, the two BHs form a bound pair. That binary then hardens  
by interactions with stars and gas. The efficiency of these processes in shrinking the binary
orbit has been much discussed in the literature, and is known as the ``final-parsec problem'',
reflecting early concerns that the binary may stall at parsec-scale separations for
more than a Hubble time, rarely reaching a regime where efficient
GW emission leads
to rapid coalescence of the system. Recent results indicate that non-axisymmetric galaxy
potentials, the abundance of stars with centrophilic orbits, and/or the presence of large amounts 
of gas are, in many cases, sufficient to shrink the orbit in less than a Hubble time (e.g.,
Perets \& Alexander 2008, Preto et al. 2011, Khan et al. 2013, Mayer 2013, Chapon et al. 2013,  
Ivanov et al. 2014, Aly et al. 2015, Vasiliev 2015; review by Colpi 2014).  
At coalescence, the emitted GWs carry away linear momentum,
so that the SMBH receives a kick velocity,  
and then oscillates about the core of its host galaxy,  
or in rare cases escapes (e.g., Campanelli et al. 2007, 
Lousto \& Zlochower 2011; review by Centrella et al. 2010).   

Key questions related to all stages of galaxy merging include
the following: (a) when does the accretion process start, 
(b) how long does it last, (c) how much matter is accreted before 
and after binary coalescence, (d) how much 
do the SMBHs grow in each phase, (e) how often are both SMBHs active, (f) how efficient are feedback
processes, (g) how efficient is the loss of angular momentum due to interactions with gas and stars,
(h) how much do the SMBHs' spins change during accretion, 
(i) how quick does the SMBBH coalesce, (j) how frequent are coalescences in the universe,
and (k) what is the amplitude of GW recoil?    

The answers to these questions are central to our understanding of the assembly
history and demography of black holes, and of galaxy formation and evolution 
across cosmic times.  
Identifying SMBBHs in all stages of their evolution is 
therefore of great interest, and  an intense search is currently ongoing.  

A variety of signatures have been used to search for and identify pairs, SMBBHs, 
and candidates (Fig. 2). Detection of the wide systems, when the two black holes
are spatially resolved from each other, is observationally most easy and robust.  
More indirect methods are in use to search for the closest SMBBH systems, no longer spatially resolved.
Semi-periodicities in lightcurves or spatial structures in radio jets, double-peaked emission lines,
and other features, have all been used
to identify candidates.  
Most methods require that both, or at least one SMBH, is active. Most difficult  
to recognize are SMBBHs at the cores of non-active galaxies. They could be widely present, and with
current methods (and with the single exception of
our own Galactic Center), we would have almost no way of detecting them. 
A recent suggestion has therefore
been to use the lightcurves of flares from tidally disrupted stars 
to search for the tell-tale signatures
of binaries in otherwise quiescent galaxies (Sect. 3.3).

This review provides a short overview of observations of, and search strategies
for, pairs and binary SMBHs at wide and close orbital separations. 
We will not cover the widest systems of
AGN pairs, in early stages of interaction, or multiple AGN in clusters of galaxies,
due to lack of space.
An accompanying review (Liu 2015, these proceedings) will elaborate in much greater depth
on theoretical aspects, and theoretical predictions of signatures of SMBBHs which have not
yet been observed, but can be used for future searches. Further, this contribution will focus
on main principles and detection methods, and a few prime representative systems. There is not enough 
space to reference all publications that have contributed to this exciting and rapidly growing field. 
Our apologies in
advance.

\section{Spatially resolved systems in single galaxies and advanced mergers}

Wide pairs of accreting SMBHs, spatially resolved, can be identified by the  
characteristic signatures of AGN activity from both BHs, 
in form of luminous (hard) X-ray emission, compact radio cores, typical optical
emission-line ratios, or IR colours.    
Only a few systems have been identified at projected separations of $r \sim$ 1 kpc or less.
In X-rays, these are NGC\,6240 (at $r = 1$ kpc; Komossa et al. 2003) and NGC\,3393 
(at $r = 150$ pc; Fabbiano et al. 2011), both
based on high-resolution \textit{Chandra} 
imaging spectroscopy. In the radio regime, two compact, variable,
flat-spectrum cores were found in 0402+379 (at $r = 7$ pc; Rodriguez et al. 2006, Burke-Spolaor
2011). In the optical band, two candidate AGN cores exist 
in SDSSJ132323.33$-$15941.9 (at $r = 0.8$ kpc; Woo et al. 2014).

The galaxy pair SDSSJ1502+1115 (Sect. 3.1) is remarkable for its overall radio
structure. It consists of two bright radio cores at 7.4 kpc separation (Fu et al. 2011b).
One of the two is further resolved into two knots of about equal brightness 
at 140 pc projected separation.   
These have been interpreted as representing
either two separate SMBHs, or else double hot spots around a single SMBH 
(Deane et al. 2014, Wrobel et al. 2014).

\section{Candidate spatially unresolved systems}

\subsection{Double-peaked emission lines}

Optical spectra of AGN are characterized by narrow and broad emission lines. If these appear double,
they may indicate the presence of two accreting SMBHs (Gaskell 1983, 1996, Zhou et al. 2004, review by
Popovic 2012). Further,
single-peaked emission lines, which are kinematically shifted with respect to their host galaxy,  
may imply the presence of a merger (Comerford et al. 2009).  

In recent years, larger samples of AGN with doubled-peaked narrow 
lines (``narrow-line double-peakers'') have
been identified thanks to large spectroscopic surveys like SDSS, AGES and LAMOST (e.g.,
Wang et al. 2009, Komossa \& Xu 2009, Liu et al. 2010, Smith et al. 2010, 
Ge et al. 2012, Comerford et al. 2013, Barrows et al. 2013, Shi et al. 2014).
A challenge when
identifying the true binary AGN among them arises from the fact, 
that several other mechanisms do exist, which
also produce double-peaked lines, but only involve a single AGN. These include the presence of
two-sided jets or outflows, rotating disks, or a {\em single} AGN which ionizes the interstellar media
of {\em two} host galaxies (e.g., Xu \& Komossa 2009). Further, double-peakers are only
expected for a short fraction of the total merger time (Yu et al. 2011,
Van Wassenhove et al. 2012, Blecha et al. 2013).
Therefore, multi-wavelength follow-up observations
are required, in order to select the binaries among the large numbers of double-peakers. Such follow-ups 
have shown that only a small fraction of them, $\sim 2\%-10\%$, 
harbor AGN pairs (e.g., Fu et al. 2011a, Fu et al. 2012, 
Shen et al. 2011, Smith et al. 2012, Comerford et al. 2012). 
One of the confirmed systems is SDSSJ1502+1115,
with two luminous radio cores at a projected spatial separation of 7.4 kpc (Fu et al.
2011b).{\footnote{Note that we list narrow double-peakers under the Section of ``unresolved
sources'', because their initial selection criterion is (in most cases) based on spatially unresolved 
emission. Follow-up imaging, when available, then often did resolve the sources. However, they usually 
consist of wider pairs of galaxies with core separations above a kpc, not fitting in
the category discussed in Section 2.}} 
   
A fraction of all quasar spectra exhibits double-peaked {\em broad} emission lines. If these are
due to two broad-line regions bound to two SMBHs orbiting each other, 
we should see the characteristic
Doppler-shifts of the emission lines reflecting the orbital motion (Gaskell 1983, Shen \& Loeb 2010). 
Broad-line double-peakers
carefully monitored in the 1980s and 90s (e.g., Halpern \& Filippenko 1988, Halpern \& Eracleous 2000) 
did not reveal the expected orbital motions,
and have been interpreted as systems with warped accretion disks around single SMBHs instead. New large
samples of broad-line double-peakers, or of systems with single, kinematically
shifted broad lines, have now been selected from SDSS (e.g., Tsalmantza et al. 2011, 
Eracleous et al. 2012, 
Decarli et al. 2013, Shen et al. 2013, Ju et al. 2013), and some binaries
may hide among them. 

Recently, Bon et al. (2012) presented a SMBBH model for the well-known, nearby, broad-line 
Seyfert galaxy NGC\,4151, based on evidence for periodic variations of the 
observed H$\alpha$ emission line, in many years of spectroscopic monitoring. 
The observations have been explained
with a sub-parsec binary with an orbital period of $\sim$16 yrs.  


\begin{figure}[t]
\begin{center}
 \includegraphics[width=9.5cm]{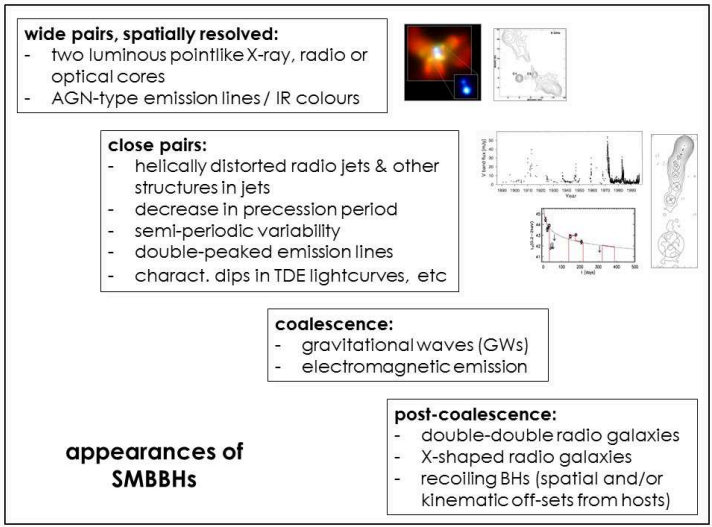}
 \caption{Signatures and detection methods of SMBH pairs and binaries.}
   \label{fig2}
\end{center}
\end{figure}
 
\subsection{Semi-periodic variability}

A number of blazars show evidence for semi-periodic optical variability, which might be linked to
the presence of a second SMBH (e.g., Sillanp{\"a\"a} et al. 1988, Raiteri et al. 2001,
Fan et al. 2002, de Paolis et al. 2002, 
Rieger 2004, Ostorero et al. 2004, 
Liu et al. 2006, Qian et al. 2007, Volvach et al. 2007, 
Xie et al. 2008, Karouzos et al. 2010, 
Kudryavtseva et al. 2011, Graham et al. 2015).  
The best studied such case is the blazar OJ287. Its optical lightcurve has been observed 
for more than a century
(thanks to photographic plate archives all around the world, 
and intense dedicated monitoring during
the last few decades)
and shows repeat outbursts every $\sim$12 yrs 
(e.g., Sillanp{\"a\"a} et al. 1988, Valtaoja et al. 2000,
Valtonen et al. 2012), 
each of which is composed of two peaks
separated by $\sim$1 yr. 
The best explored SMBBH model consists of a secondary BH in a precessing orbit,
which impacts a warped accretion disk around the primary twice each orbit. 
Precise timing of the past optical peaks then allows to derive the orbital
parameters of the system. Calculating Keplerian orbits with post-Newtonian corrections,
Valtonen (2007) presented an orbital solution with a primary mass 
of $\sim$2\,10$^{10}$ M$_{\odot}$, a mass ratio $\sim$0.01, an eccentricity of $\epsilon=0.66$,
and a semi-major axis of 0.045 pc. They also reported tentative evidence for an orbital
shrinkage due to emission of GWs, of order $\Delta T_{\rm GW} \approx 0.01$ yr/period 
(Valtonen et al. 2008).    
The next optical maximum is expected in a few years, allowing to test new predictions
of all recent models for OJ287. 

\subsection{Dips in TDE lightcurves}

SMBBHs imprint their presence on the outburst lightcurves of stellar tidal disruption events (TDEs). 
The secondary temporary interrupts the accretion stream on the primary, causing characteristic
deep dips in the decline lightcurves (Liu et al. 2009). This signature has been observed in the
lightcurve of the TDE from SDSSJ120136.02+300305.5, which is well modelled with a binary
of mass ratio $q \sim 0.1$  
at $\sim$0.6 mpc spatial separation (Liu et al. 2014). In the future, SMBBHs with TDEs may also 
be recognized by reprocessed emission lines which will show
a tilted response function due to the off-centre location of one of the SMBHs (Brem et al. 2014).

\subsection{Structures in radio jets} 

Several blazar radio jets show semi-periodic deviations from a straight
line, and/or some other unusual structures. One way to explain these observations 
is involving the presence of a binary SMBH, which
causes either (1) a modulation due to orbital motion of the jet-emitting BH around the primary BH,
or (2) jet precession (e.g., Begelman et al. 1980, Roos 1988, Hardee et al. 1994,
Britzen et al. 2001){\footnote{See, e.g., Britzen et al. 2010, Lobanov \& Roland 2005,
and Godfrey et al. 
2012 (and references therein), for a discussion including alternative scenarios such as disk
oscillations or Kelvin-Helmholtz instabilities)}}. 
If the jet precession is caused by a binary, 
then a prediction of this scenario is 
the acceleration of jet precession, observable
on long timescales (Liu \& Chen 2007).

Radio interferometry has provided us with the highest-resolution observations of jets 
over decades. Here, we would like to mention three representative candidate SMBBH systems.
These are among the well-studied systems, but there is a 
number of others which would deserve
mentioning, and are not due to lack of space.
The quasar S5\,1928+738 has long been suspected to harbor a SMBBH 
(Hummel et al. 1992, Roos et al. 1993, Murphy et al. 2003, see also Roland et al. 2014). 
Kun et al. (2014), analyzing 20 yrs of VLBI data, 
presented evidence that the jet-emitting SMBH  is actually spinning. 
Their orbital modelling implies a binary separation of $\sim$10 mpc,
and an orbital period of $\sim$5 yrs.  
The helical distortions of the jet of the BL Lac object Mrk 501 have been interpreted 
with a SMBBH model by Conway \& Wrobel (1995) and Villata \& Raiteri (1999).   
A SMBBH scenario was also involved in order
to explain evidence for semi-periodic variability of this source (Rieger \& Manheim 2000;
see also de Paolis et al. 2002, R{\"o}dig et al. 2009).    
Lobanov \& Roland (2005) presented a SMBBH model at $\sim$0.3 pc
separation for the quasar 3C345, which can reproduce both,
its optical and radio variability, and the morphology and kinematics of the parsec-scale jet.   

The most powerful method to date of {\em spatially} resolving the orbit of a jet-emitting BH in
a compact binary is phase-referencing of VLBI radio data. 
Using that technique, Sudou et al. (2003) reported evidence for systematic changes in radio
position, which they interpreted as  
orbital motion of the radio core of 3C66B with a period of 1.05 yr. 
Part of the possible orbital solutions
could be excluded based on current pulsar timing constraints (Jenet et al. 2004), while the rest
remains a possibility (Iguchi et al. 2010). 

Future phase-reference measurements of this and other systems, along with  
simulations of the jet base, core-shift measurements, and studies of transverse
motions will provide us with strong tests of the
SMBBH model.

\section{Post-coalescence candidates}

Certain signatures of compact and coalescing binaries remain imprinted on their large-scale
environment, and can therefore be recovered from multi-wavelength observations long after
the actual coalescence. For instance, accretion temporarily interrupts in compact binaries, 
because of the fast
orbital shrinkage due to GW emission, dominating over viscous processes,
so that the inner disk no longer catches up (e.g., Liu
et al. 2003, Milosavljevic \& Phinney 2005, Farris et al. 2015). 
If these systems launch radio jets, jet formation will be temporarily
interrupted, too, and this may explain the presence of double-double radio galaxies (Liu et al. 2003).
If the hole's spin direction changes after coalescence, the jet will be launched in a new direction,
and this is one possibility to account for the structure of X-shaped 
radio galaxies (Merritt \& Ekers 2002; see Gopal-Krishna et al. 2012 for a recent overview; 
see also Mezcua et al. 2012).   
If the newly formed single SMBH receives a significant kick velocity after coalescence,
it will appear spatially or kinematically off-set from its host galaxy, and
several candidate recoiling SMBHs have emerged in recent years (review by Komossa 2012).  
Further, it has been suggested that the central stellar light deficits observed in some ellipticals
and bulges were created by SMBBHs which had shrunk their orbits by slingshot ejection
of stars, consistent with recent observations (e.g., Dullo \& Graham 2014).    

\section{Future missions and searches}

A number of ongoing and future missions and surveys will be sensitive to SMBBHs
in all stages of evolution.  For instance, space VLBI and mm VLBI at the shortest
wavelengths feasible will provide us with
the highest spatial resolution (e.g., Fish et al. 2013,
Tilanus et al. 2014), while the \textit{Square Kilometer Array} (SKA) will provide 
high sensitivity (e.g., Deane et al. 2015).   
SKA and other current and upcoming PTA (pulsar timing array) experiments will  
detect the gravitational wave signatures of the most massive coalescing SMBBHs using   
pulsar timing (e.g., Lazio 2013, Hobbs 2013, Kramer \& Champion 2013, Sesana 2015). 
Future high-sensitivity X-ray observatories may kinematically resolve
binary effects on the iron line profile from one or two disks 
(Yu \& Lu 2001, McKernan et al. 2013, 	
Jovanovic et al. 2014), while 
optical integral field spectroscopy may reveal the kinematic 
signature of the inspiral phase (Meiron \& Laor 2013).   
 
Further breakthroughs in the field are expected once space-based 
gravitational-wave interferometers
are in operation, providing measurements of 
coalescence rates, SMBH masses and spins 
(e.g., Babak et al. 2011, review by Barausse et al. 2015). 
eLISA is currently scheduled for launch around 2030. 
Electromagnetic counterparts to GWs (e.g., Schutz 1986) from coalescing SMBBHs,
and characteristic signals before or after coalescence, 
may appear as transients in current or future  transient surveys 
(reviews by Schnittman 2011, Haiman 2012).


\section{Tidal disruption of stars by supermassive black holes}

The tidal disruption, and subsequent accretion, of a star by
a supermassive black hole produces a luminous flare of electromagnetic
radiation (e.g., Rees 1990, Luminet 1985).
A star 
is  disrupted, once the tidal forces of the hole  
exceed the self-gravity of the star
(Hills 1975).
The distance at which this happens, the tidal radius, is given by
\begin{equation}
r_{\rm t} \simeq 7\,10^{12}\,{\bigg{(}}{M_{\rm BH}\over {10^{6} M_\odot}}{\bigg{)}}^{1 \over 3}
    {\bigg{(}}{M_{\rm *}\over M_\odot}{\bigg{)}}^{-{1 \over 3}} {r_* \over r_\odot}~{\rm cm}\,. 
\end{equation}
A fraction of the stellar material
will be on unbound orbits and escape, while the rest will
eventually be accreted (Fig. 3).  The events appear as luminous transients
with peak in the UV or soft X-rays, declining on the timescale of months to years
(e.g., Rees 1990, Evans \& Kochanek 1989).  Recent state of the art modelling 
has addressed the different stages of TDE evolution under various conditions
(e.g., Lodato et al. 2009, Brassart \& Luminet 2010, Strubbe \& Quataert 2011, 
Lodato \& Rossi 2011, Cheng et al. 2012, 
Kesden 2012, Guillochon \& Ramirez-Ruiz 2013, Hayasaki et al. 2013,
Dai \& Blandford 2013, Cheng \& Bogdanovic 2014, 
Shiokawa et al. 2015; and references therein). 
If the doomed star is compact (e.g., a white dwarf), 
partial disruption will produce an electromagnetic {\em and} a GW signal 
(review by Amaro-Seoane et al. 2007).   

Observing TDEs, out to large cosmic distances, 
provides us with a variety of new astrophysical tools and applications,
related to fundamental physics or astrophysics, including 
studying precession effects in the Kerr metric, measuring BH spin,
observing other relativistic effects (at 10$^8$ M$_{\odot}$, the tidal
radius is on the order of the Schwarzschild radius), 
probing accretion physics under extreme conditions and near $L_{\rm edd}$,
understanding the physics of jet formation and early evolution, 
reverberation-mapping the gaseous core environment via its emission-line response,
searching for a population of (so far elusive) intermediate mass BHs,
detecting supermassive binary BHs at the cores of quiescent galaxies (from TDE
lightcurves; Sect. 3.3), probing stellar kinematics on spatial
scales which cannot be resolved directly (via disruption rates in different
types of galaxies), or spotting recoiling BHs by off-nuclear TDEs.


\begin{figure}[t]
\begin{center}
\includegraphics[width=10.5cm]{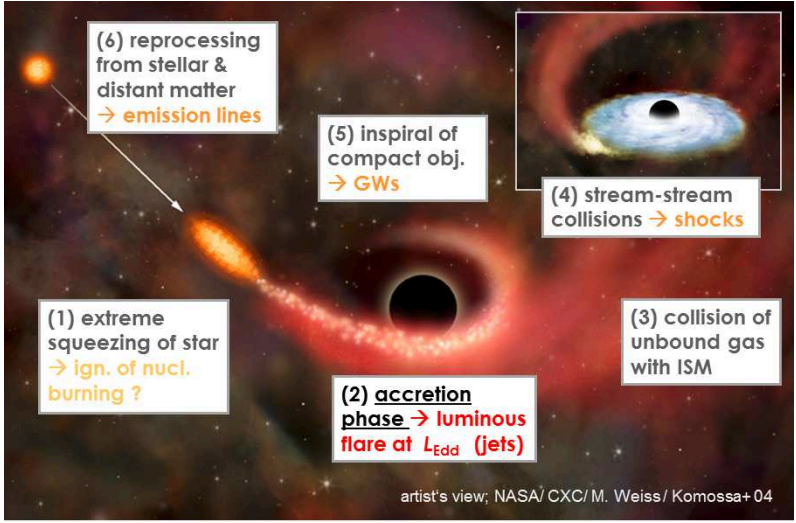} 
 \caption{Evolution of stellar tidal disruption events and sources of radiation. In
most cases, the accretion phase is the most luminous electromagnetic phase. 
}
   \label{fig3}
\end{center}
\end{figure}

\section{Multi-wavelength observations}

A key signpost of TDEs is their luminous, transient high-energy emission, peaking in the 
UV or soft X-rays, arising from the accretion of the stellar material.  
First events from quiescent galaxies have been identified in the course of the ROSAT
all sky survey, which was ideal for detection because of its repeat coverage of almost
the whole sky, and its high sensitivity in the soft X-ray band (0.1-2.4 keV). 
Events appeared as luminous transients, reaching peak luminosities up to $> 10^{43-44}$ erg/s 
just in soft X-rays (e.g., Bade et al. 1996, Komossa \& Bade 1999, 
Grupe et al. 1999). They then faded away
by factors larger than a few thousand (Halpern et al. 2004, Komossa et al. 2004),
their initially supersoft X-ray spectra ($kT \sim 0.04-0.1$ keV) showed a hardening with time,
and optical spectroscopy of the host galaxies revealed little or no 
activity at all (review by Komossa 2002). 
Two of these events, NGC\,5905 and RXJ1242$-$1119, continue to be the best-monitored
events in terms of their {\em long-term}
X-ray lightcurves, spanning time intervals of more than a decade (Komossa et al. 2004,
Halpern et al. 2004). All the event properties agree very well with order-of-magnitude
predictions from tidal disruption theory (e.g., Rees 1988, 1990).
More recently, in X-rays similar events have been found with 
\textit{Chandra} and \textit{XMM-Newton} (e.g., Esquej et
al. 2008, Maksym et al. 2010, Lin et al. 2011, Saxton
et al. 2012, Nikolajuk \& Walter 2013, Maksym et al. 2013, Donato et al. 2014), some of them with 
well-covered lightcurves during the first few years. 
Events were also found at longer wavelengths, in the UV and optical (e.g.,
Gezari et al. 2006, Komossa et al. 2008,   
van Velzen et al. 2011, Cenko et al. 2012a,
Gezari et al. 2012, Chornock et al. 2014), some of them caught before their peak
(see Komossa 2012 for a more extended review of multi-$\lambda$ observations).  
Several estimates of TDE rates are all on the order of $10^{-4} - 10^{-5}$/yr/galaxy (e.g., 
Donley et al. 2002, 
Esquej et al. 2008, Maksym et al. 2010, Wang et al. 2012) 
and agree well with  
theoretical predictions (e.g., Brockamp et al. 2011).  

\section{Emission-line transients}

TDEs which occur in gas-rich galaxies will provide us with
a powerful new tool of performing reverberation mapping of the
cores of these galaxies.
As the luminous electromagnetic radiation
travels across the galaxy core, it will photoionize
any circum-nuclear material (including the tidal debris itself) and is reprocessed into
line radiation.
Recently, SDSS and other surveys have enabled the discovery of 
several well-observed cases of transient
optical emission lines, of a kind not observed before, and arising from otherwise 
quiescent galaxies{\footnote{These emission-line transients are markedly different from
the mild line variability seen in AGN, with two exceptions:  
(1) The AGN IC\,3599, which underwent a high-amplitude
X-ray outburst accompanied by a strong increase in its optical emission lines 
(Brandt et al. 1995, Grupe
et al. 1995, Komossa \& Bade 1999), and (2) the AGN NGC\,1097, which 
shows strong, broad, double-peaked
Balmer lines which emerged abruptly  
(e.g., Storchi-Bergmann et al. 1995).     
The underlying mechanism remains unknown, but 
high-amplitude Narrow-line
Seyfert 1 variability (only IC\,3599), variants of accretion-disk instabilities, or a TDE have all been 
considered.}}:  
All of them exhibit bright, broad, fading emission from Helium and/or Hydrogen 
(Komossa et al. 2008, 2009,
Wang et al. 2011, 2012, Gezari et al. 2012, Gaskell \& Rojas Lobos 2014, Holoien et al. 2014,
Arcavi et al. 2014), while
some of them show transient super-strong iron coronal lines in 
addition, up to ionization stages of Fe$^{13+}$
(Komossa et al. 2008, 2009, Wang et al. 2011, 2012).

\section{Jetted TDEs}

The possibility that TDEs launch radio jets, came up 
with the detection of the first few X-ray TDEs with ROSAT. Dedicated follow-ups of NGC\,5905
did not detect any radio emission from a jet, however (Komossa 2002). 

Two events recently discovered with \textit{Swift}, Swift\,J1644+57
and Swift\,J2058.4+0516, differ from previous TDEs,
in the sense that they had much harder X-ray spectra,
were accompanied by strong (beamed) radio emission,
and exhibit some other remarkable properties
(e.g., Burrows et al. 2011, Bloom et al. 2011,
Zauderer et al. 2011, 2013, Levan et al. 2011, Cenko et al. 2012b).
Swift\,J1644+57 was detected with \textit{Swift} BAT in 2011.  
Its (isotropic) peak luminosity
exceeded $10^{48}$ erg/s. The X-ray lightcurve shows a
general downward trend,
on which rapid, high-amplitude variability is superposed, 
as fast as 100s. After $\sim$1.5 yrs, the X-rays suddenly
dropped by a large factor, and have remained faint so far.
The host galaxy at redshift $z=0.35$ does
not show signs of permanent optical AGN activity.
The event is accompanied
by unresolved and variable radio emission, which has been interpreted
as the rapid onset of a powerful jet after stellar tidal disruption.
The event has motivated a large number of follow-ups and theoretical
studies (review by Komossa 2015, in prep.), with an emphasis on the
question of jet launching under TDE conditions, and the role of magnetic fields
(e.g., Tchekhovskoy et al. 2014).  

These and future observations of jetted TDEs  provide us with a completely new probe of
the early phases of jet formation and evolution in an otherwise quiescent environment
without past radio-AGN activity.

\section{Future missions and surveys}

TDEs will be detected in large numbers with future sky surveys, including in the radio
with SKA (Donnarumma et al. 2015), in the optical with LSST (Gezari et al. 2009), 
in hard X-rays with LOFT (Rossi et al. 2015), and in soft X-rays with the proposed mission
{\em Einstein Probe} (Yuan et al. 2015).  Well-covered 
lightcurves will enable a wealth of new science, and X-rays will be sensitive to relativistic
effects (Sect. 6).

SK would like to thank ISSI/Bern for supporting and hosting two 
workshops on ``Unveiling multiple AGN
activity in galaxy mergers'',
and the participants for many stimulating discussions. SK would also like to thank NAOC Beijing
for their great hospitality and support over many years. 
Many thanks to S. Britzen, T. Krichbaum,
A. Lobanov, and E. Ros for a critical reading of the manuscript and very useful comments.

\end{document}